\documentclass[aps,prl,twocolumn,nofootinbib,groupedaddress,amsfonts,floatfix]{revtex4-1}

\usepackage{graphicx,amsmath,amssymb,amstext}
\usepackage{amssymb,amsbsy,amsfonts,amsthm,color}
\usepackage[export]{adjustbox}
\pdfoutput=1
\usepackage{booktabs}
\usepackage[english]{babel}
\usepackage{amsmath,amssymb,amsbsy,amstext, amsthm, simplewick}
\usepackage{wrapfig}
\usepackage{graphicx}
\usepackage{amsfonts}
\usepackage{amssymb}
\usepackage{upgreek}
 \usepackage{exscale,relsize}
 \usepackage[makeroom]{cancel}
\usepackage{soul}
\usepackage{bbold}
\usepackage{enumerate}
\usepackage{braket}

\RequirePackage{color}
\usepackage{colortbl}
\definecolor{rp}{cmyk}{0.2, 1, 0.6, 0}
\definecolor{green2}{cmyk}{0, 1, 0.5, 0}
\definecolor{lightgreen}{cmyk}{0.2, 0, 0.2, 0.2}
\definecolor{lightgray}{cmyk}{0.1,0.2,0,0.1}
\definecolor{lightgray2}{cmyk}{0.4,0.4,0,0.8}
\definecolor{black}{cmyk}{1.0,1.0,1.0,1.0}

\allowdisplaybreaks[1]

\DeclareMathAlphabet{\mathpzc}{OT1}{pzc}{m}{it}

\usepackage{colortbl}
\definecolor{lightgreen}{cmyk}{0.2, 0, 0.2, 0.2}
\definecolor{lightgray}{cmyk}{0.1,0.2,0,0.1}
\definecolor{lightgray2}{cmyk}{0.1,0.1,0,0.1}

\makeatletter
\newlength{\apb@width}
\newcommand{\autoparbox}[2][c]{\settowidth{\apb@width}{#2}\parbox[#1]{\apb@width}{#2}}

\makeatother

\newcommand{\lsim}{\mathrel{\hbox{\rlap{\lower.55ex\hbox{$\sim$}} \kern-.3em \raise.4ex \hbox{$<$}}}}
\newcommand{\gsim}{\mathrel{\hbox{\rlap{\lower.55ex\hbox{$\sim$}} \kern-.3em \raise.4ex \hbox{$>$}}}}

\newcommand{\mpl}{m_{\mbox{\tiny{Pl}}}}
\newcommand{\Beq}{\begin{equation}\begin{aligned}}
\newcommand{\Eeq}{\end{aligned}\end{equation}}

\newcommand{\bk}{{\textbf{\textit{k}}}}
\newcommand{\bx}{{\textbf{\textit{x}}}}

\begin{document}

\title{The Equation of State and Duration to Radiation Domination After Inflation}

\author{Kaloian D. Lozanov${}^1$}

\author{Mustafa A. Amin${}^2$}

\affiliation{${}^1$Institute of Astronomy, University of Cambridge, CB3 0HA Cambridge, U.K.}

\affiliation{${}^2$Physics \& Astronomy Department, Rice University, Houston, Texas 77005-1827, U.S.A.}

\date{\today}
\begin{abstract}	
We calculate the equation of state after inflation and provide an upper bound on the duration before radiation domination by taking the nonlinear dynamics of the fragmented inflaton field into account. A broad class of single-field inflationary models with observationally consistent flattening of the potential at a scale $M$ away from the origin, $V(\phi)\propto |\phi|^{2n}$ near the origin, and where the couplings to other fields are ignored are included in our analysis. We find that the equation of state parameter $w\rightarrow 0$ for $n=1$ and $w\rightarrow 1/3$ (after sufficient time) for $n\gtrsim 1$. We calculate how the number of $e$-folds to radiation domination depends on both $n$ and $M$ when $M\sim \mpl$, whereas when $M\ll \mpl$, we find that the duration to radiation domination is negligible. Our results are explained in terms of a linear instability analysis in an expanding universe, scaling arguments, and are supported by 3+1 dimensional lattice simulations. We show how our work significantly reduces the uncertainty in inflationary observables, even after including couplings to additional light fields.
\end{abstract}
\maketitle
\noindent {\em Introduction} --- Inflationary cosmology provides a consistent framework for calculating the initial conditions responsible for the observed temperature fluctuations in the cosmic microwave background \cite{Ade:2015lrj}. However, there is a gap in our understanding of how inflation ends and ultimately leads to a radiation-dominated, thermal universe before the production of light elements. The poorly constrained post-inflationary equation of state of the universe and the duration before radiation domination influence the interpretation of inflationary observables and the reheating temperature \cite{Liddle:2003as,Adshead:2010mc,Creminelli:2014oaa,Dai:2014jja,Martin:2014nya,Munoz:2014eqa,Cook:2015vqa,Ellis:2015pla,Ueno:2016dim,Eshaghi:2016kne}; they affect predictions for baryogenesis and primordial relics \cite{Giudice:1999yt,Hertzberg:2013jba,Kane:2015jia}.

In this {\it Letter} we calculate the equation of state parameter $w$ soon after the end of inflation by accounting for the full nonlinear dynamics of the inflaton field using 3+1 dimensional lattice simulations. Using our results, we can calculate an upper bound on the duration to radiation domination. This bound significantly reduces the uncertainty in the interpretation and calculation of inflationary and post-inflationary observables.


The equation of state for oscillating {\it homogeneous} condensates in an expanding universe has been well understood since the 1980's \cite{Turner:1983he}; however, general results for the cases where the scalar field undergoes significant fragmentation are not easily found in the literature. Detailed earlier works on the equation of state including nonlinear dynamics certainly exist, e.g. \cite{Podolsky:2005bw}, but are usually limited to quadratic and quartic inflaton potentials coupled to light fields. We allow for general shapes of the inflaton potential, ignore couplings to other light fields in our simulations but include them in the bounds on the duration to radiation domination.

\begin{wrapfigure}{R}{0.26\textwidth}
\vspace{-0.75cm}
\begin{center}
	\includegraphics[width=0.262\textwidth]{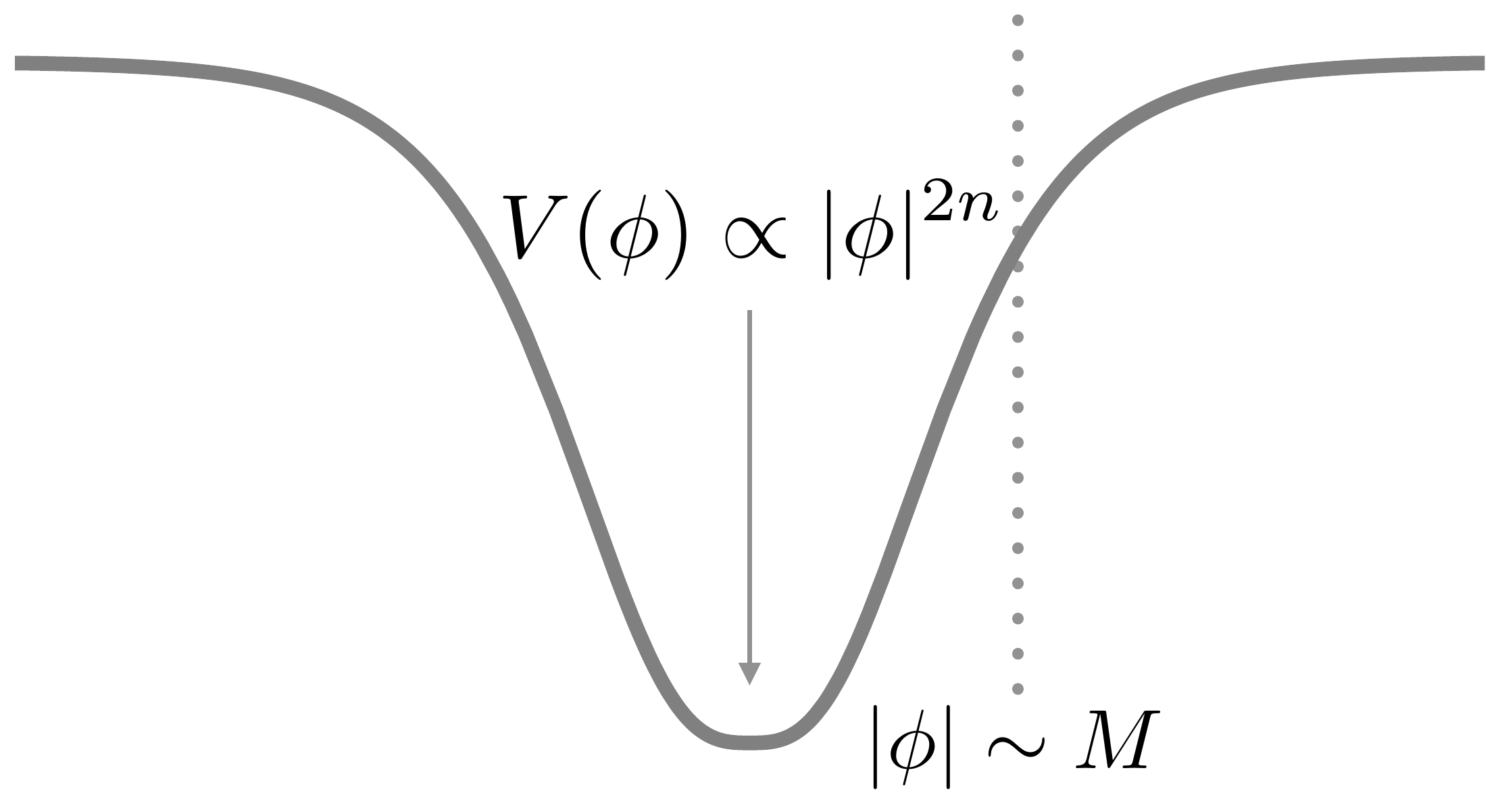}
	\caption{}
	\label{}
\end{center}
\vspace{-0.75cm}
\end{wrapfigure}
\noindent {\em Inflaton Potential} --- We study the post-inflationary expansion history in minimally coupled, single-field models of inflation with potentials of the form $V(\phi)\propto |\phi|^{2n}$ near the origin and appropriately flattened away from it (to be consistent with observations \cite{Ade:2015lrj}). For our purposes, only two features of the potential are relevant: the scale $M$ where the potential starts flattening and the power $n$ of the potential near the minimum. For concreteness, we parametrize the inflationary potentials as $V(\phi)=\Lambda^4\tanh^{2n}\left({|\phi|}/{M}\right)$, where $M=\sqrt{6\alpha}\mpl$ based on the $\alpha$-attractors models of inflation \cite{Kallosh:2013xya,Carrasco:2015pla,Kallosh:2016gqp}. We expect our results to be independent of the details of this parametrization and equally applicable to Monodromy type models \cite{Silverstein:2008sg,McAllister:2014mpa}. We also do not expect qualitative changes when we make the potential asymmetric. 
Typical models have $M\!\sim\! \mpl$; however, we also allow for $M\ll \mpl$. To avoid numerical trouble from discontinuous higher derivatives of the potential, we assume $n\ge 1$ (not necessarily an integer). 

\begin{figure*}[t!] 
   \centering
   \hspace{-0.15in}   
   \raisebox{-0.5\height}{\includegraphics[height=1.33in]{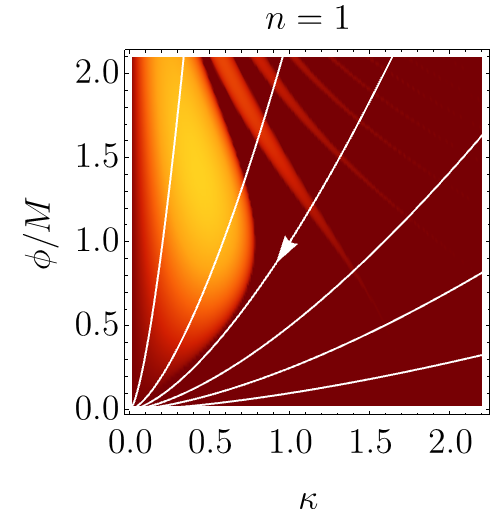}} 
   \hspace{-0.1in}
   \raisebox{-0.41\height}{\includegraphics[width=0.4in]{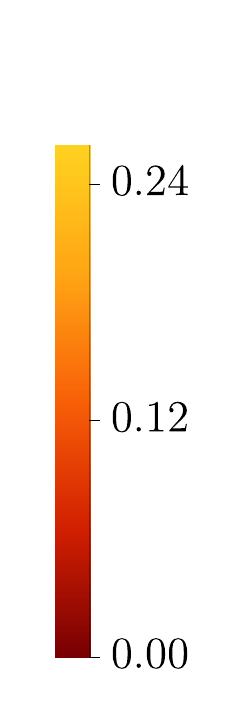}}
   \hspace{-0.11in}
   \raisebox{-0.5\height}{\includegraphics[height=1.33in]{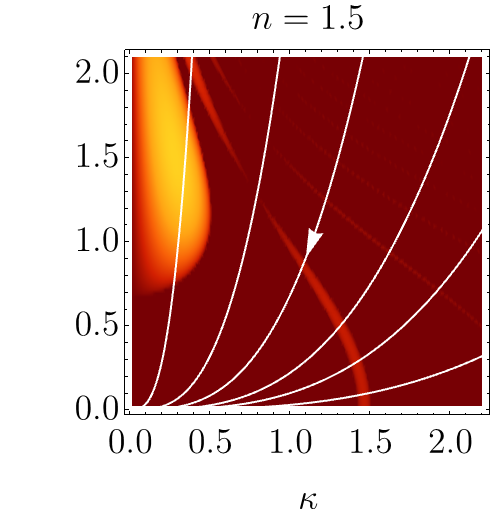}}
   \hspace{-0.1in}
   \raisebox{-0.41\height}{\includegraphics[width=0.4in]{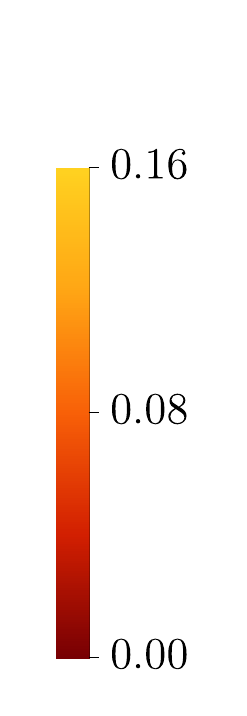}}
   \hspace{-0.11in}
   \raisebox{-0.5\height}{\includegraphics[height=1.33in]{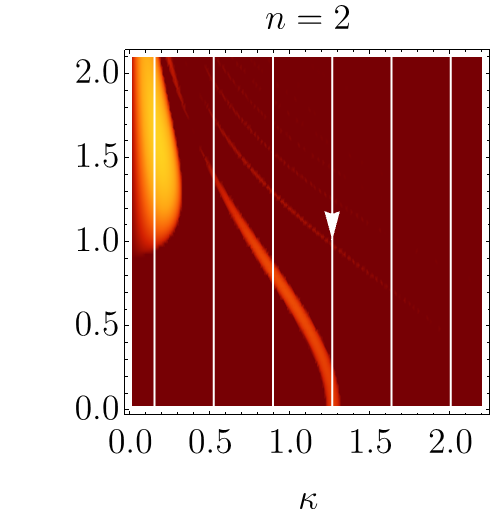}}
   \hspace{-0.1in}
   \raisebox{-0.41\height}{\includegraphics[width=0.4in]{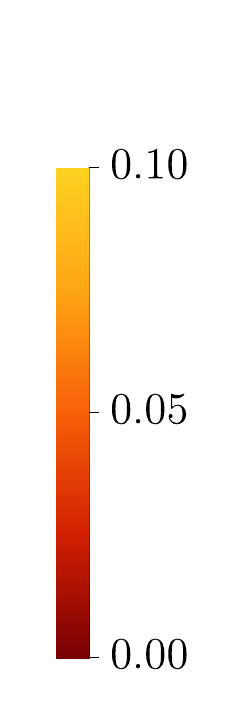}} 
   \hspace{-0.11in}
   \raisebox{-0.5\height}{\includegraphics[height=1.33in]{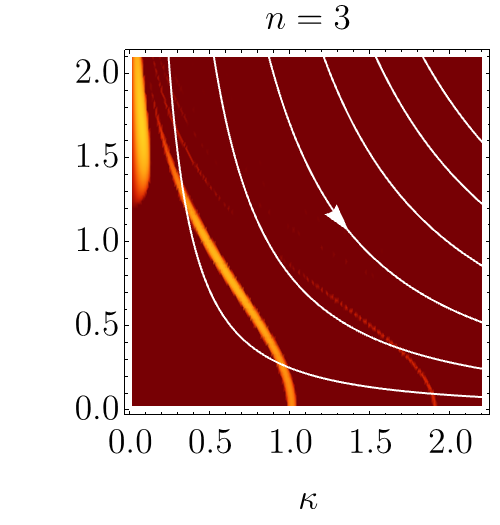}} 
   \hspace{-0.1in}
   \raisebox{-0.41\height}{\includegraphics[width=0.4in]{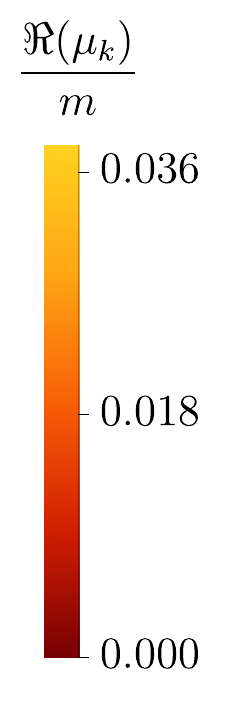}}\\
   \caption{The instability bands and the magnitude of the Floquet exponent (in units of the field dependent effective mass $m(\bar{\phi})$) are shown as functions of the oscillating condensate amplitude and the dimensionless physical wavenumber $\kappa = k/am$.  The white lines indicate how a given co-moving wavenumber passes through the instability bands as the universe expands.}
   \label{fig:Floq}
\end{figure*}
\noindent {\em Linear Instability Analysis} --- At the end of inflation, the homogeneous inflaton condensate $\bar{\phi}$ starts oscillating around the minimum of its potential. In the presence of any perturbations, such homogeneous oscillations are unstable: they lead to a rapid growth in field perturbations $\delta\phi(t,\bx)$, or equivalently, to non-adiabatic particle production \cite{Kofman:1994rk,Shtanov:1994ce,Kofman1997,Amin2014}.

A useful way of characterizing the efficiency of particle production is as follows. First, let us ignore expansion. Floquet theory tells us that the general solution for the field perturbations in Fourier space is of the form
$\delta\phi_\bk\propto\exp(\pm\mu_{k}t),$ where $\mu_k$ is the Floquet exponent. If $\Re(\mu_k)\neq0$, then there is an `unstable' solution growing exponentially with time. In general, any nonlinearity in $V(\phi)$ will lead to resonant particle production. The real part of the Floquet exponent, which characterizes the particle production rate, is shown in Fig. \ref{fig:Floq} as a function of the amplitude of the oscillating condenstate and the physical wavenumber $\kappa\equiv k/am$ (with $a=1$). Note that we have expressed $k$ and $\mu_k$ in units of a field/time dependent effective mass scale:
$
m^2\equiv2n\Lambda^2\left({\Lambda}/{M}\right)^{\!2}\left({\bar{\phi}}/{M}\right)^{\!2(n-1)}\,.
$
This effective mass scale $m^2\approx\partial_{\bar{\phi}}V/\bar{\phi}$ when $\bar{\phi}\ll M$ and is what sets the period of $\bar{\phi}$.

The expansion of the universe can now be incorporated qualitatively. The amplitude of the inflaton field oscillating in $V\propto|\phi|^{2n}$ decays as $\bar{\phi}\propto a^{-3/(n+1)}$, and the dimensionless wavenumber scales as $\kappa\propto a^{-2(2-n)/(1+n)}$. Hence a given Fourier mode flows through a number of Floquet bands as shown in Fig. \ref{fig:Floq}. Heuristically, the mode will grow if the expansion rate $H$ is much less than $|\Re(\mu_{k})|$. Strong resonance occurs for $|\Re(\mu_{k})|/H\gsim \mathcal{O}[10]$. For the lowest-$k$ band ($k/am$ near $0$):
\Beq
\left[{|\Re(\mu_{k})|}/{H}\right]^{0}_{\text{max}}={f(n)}({\mpl}/{M}),
\Eeq
where $f(n)\lesssim \mathcal{O}[1]$ with a very weak dependence on $n$ for moderate values of $n$. It is $M/\mpl$ that controls whether there is efficient self-resonance at low wave-numbers. In particular, for $M\lesssim 2.5\times10^{-2}\mpl$, the fluctuations grow rapidly and become energetically comparable to the homogeneous condensate. They backreact on the condensate, leading to its complete fragmentation. 
\begin{figure*}[t] 
\centering
   \hspace{-0.15in}
   \includegraphics[height=1.30in]{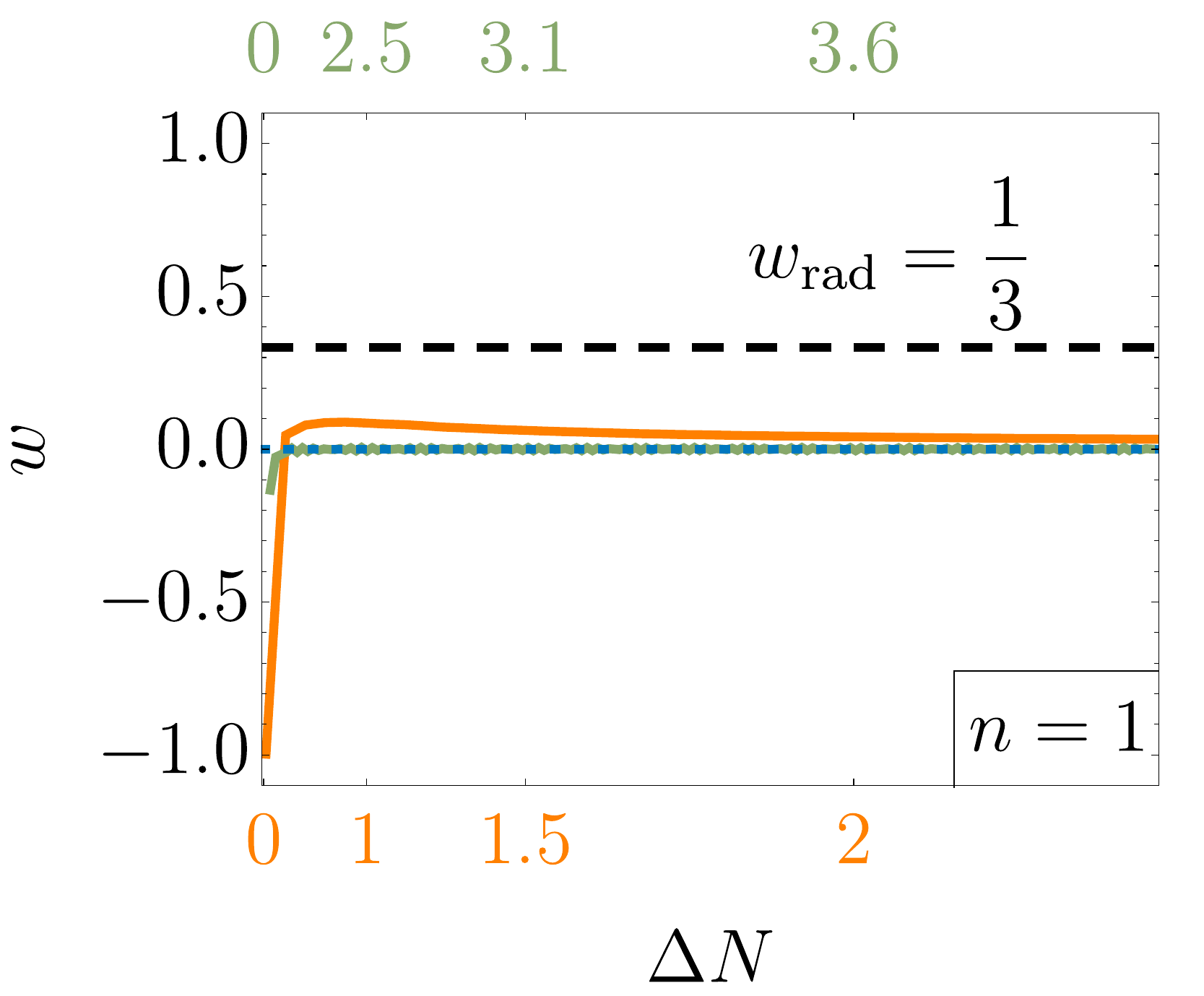} 
   \includegraphics[height=1.30in]{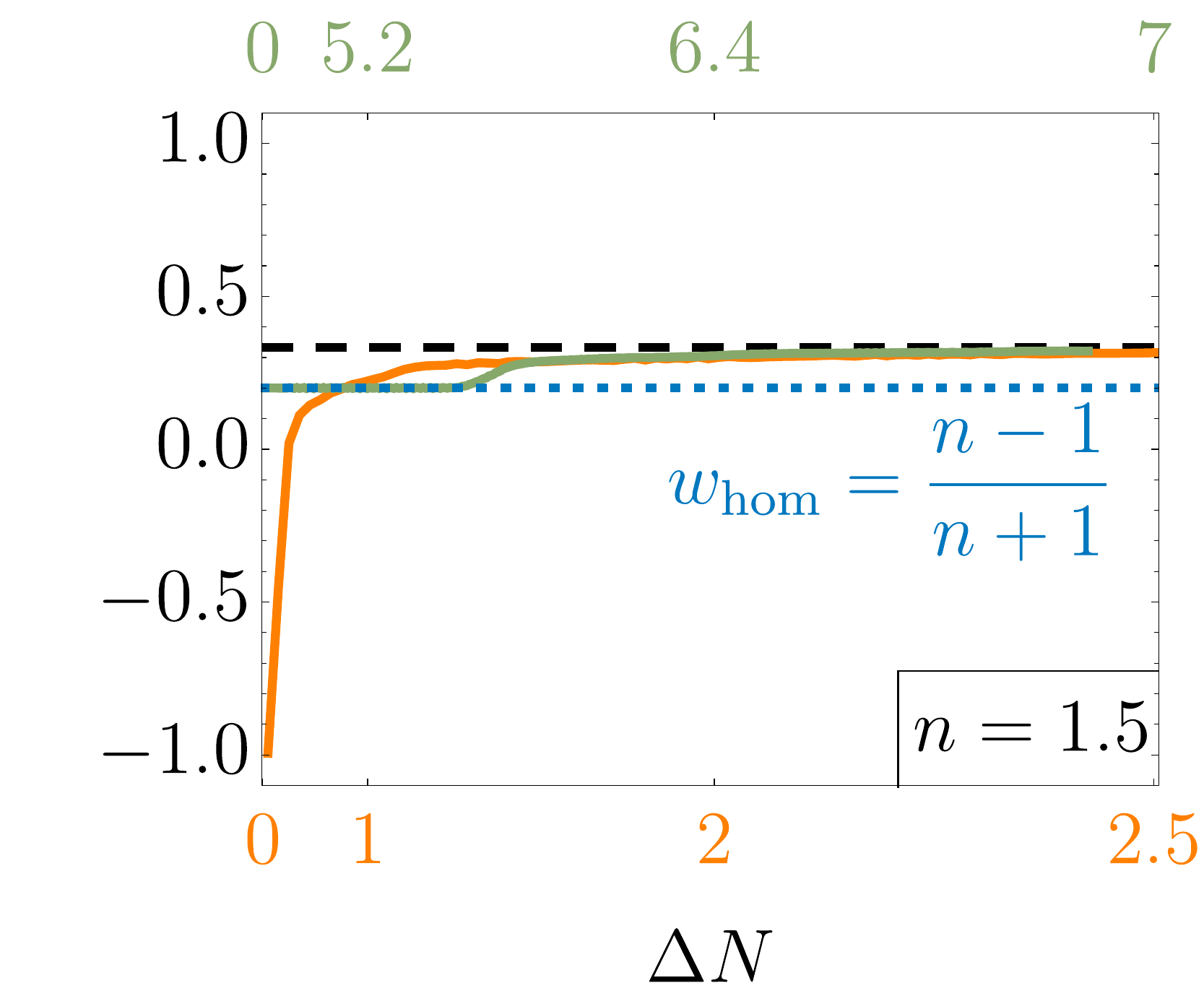} 
   \includegraphics[height=1.30in]{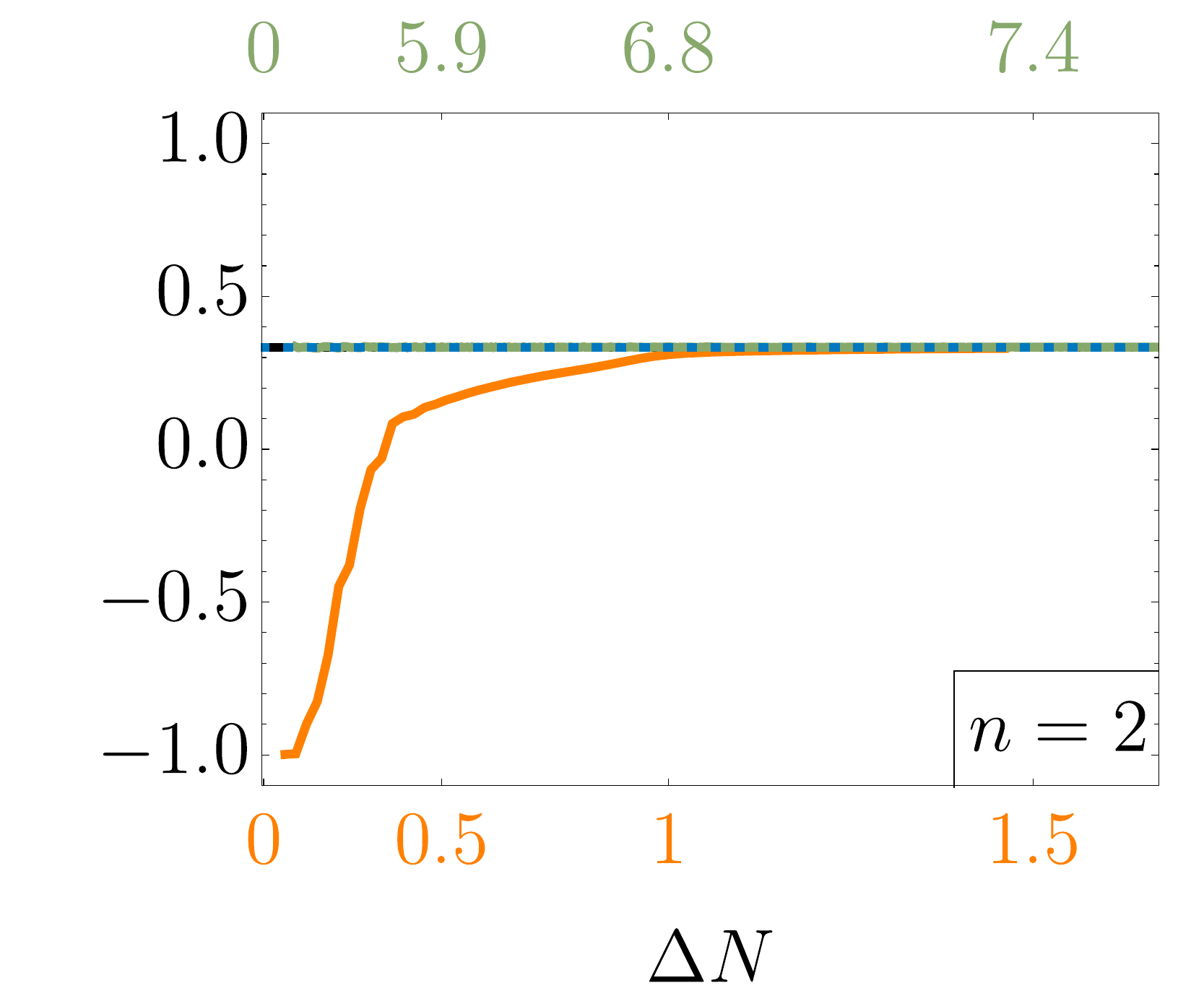} 
   \includegraphics[height=1.30in]{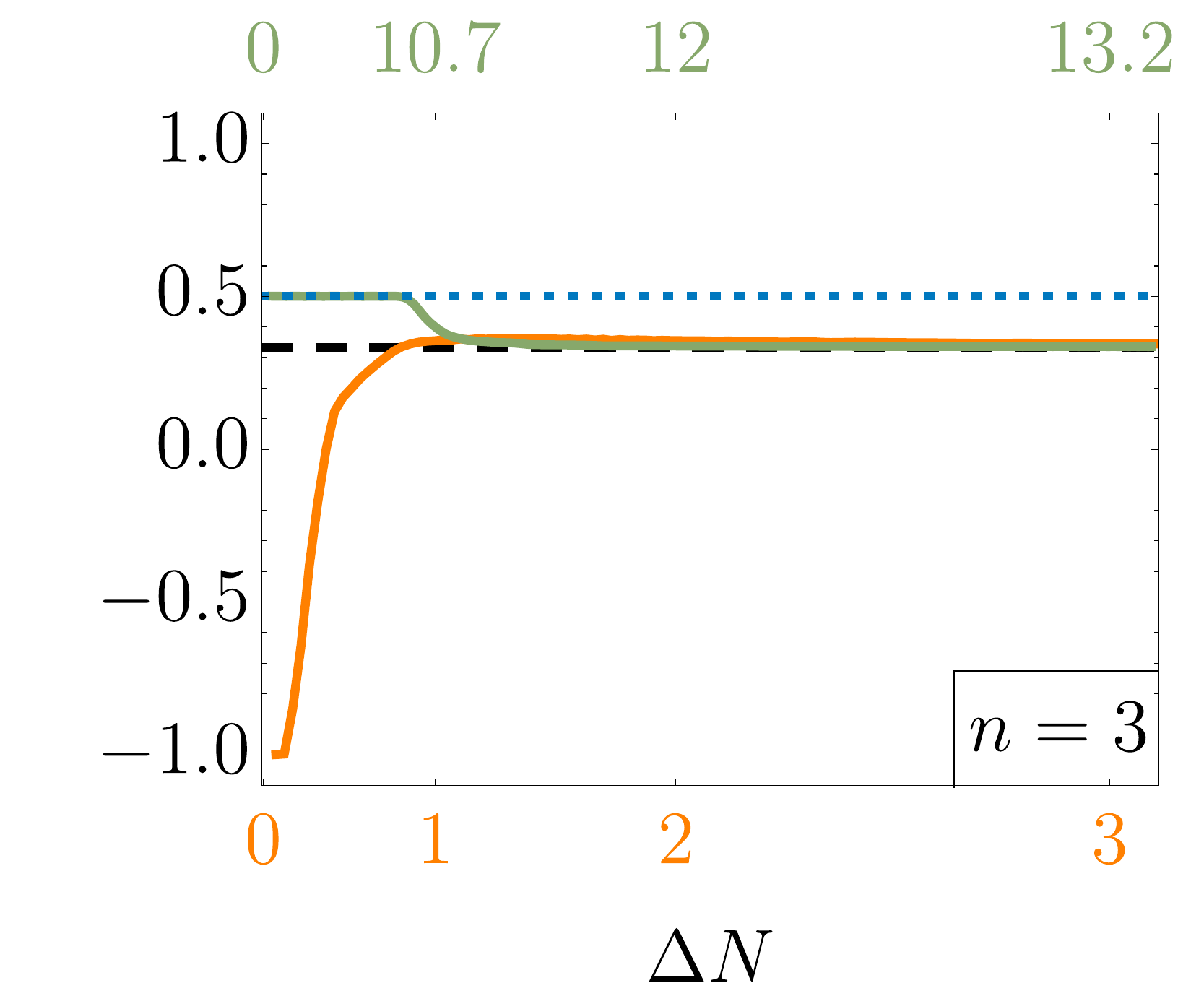} 
   \caption{The equation of state parameter obtained from the numerical simulations is shown for different values of $n$ and $M$. The orange curve and green curves correspond to initially efficient ($M\approx7.75\times 10^{-3}\mpl$) and inefficient resonance ($M\approx 2.45\mpl$), with $M\sim 2.5\times 10^{-2}\mpl$ separating the two regimes. The horizontal axes show the number of $e$-folds after the end of inflation for efficient (orange, bottom axis) and inefficient (green, top axis) resonance. The dashed line is drawn at $w=1/3$ and the dotted line denotes the homogeneous equation of state.}
   \label{fig:EoS}
\end{figure*}

When the initial fragmentation is inefficient ($M\gtrsim 2.5\times10^{-2}\mpl$), the higher order instability bands can play an important role.  Compared to the band near $k=0$, the bands at higher $k$ are narrower, and $\Re\left(\mu_k\right)$ is typically smaller. However, these narrow bands can lead to fragmentation of the condensate at late times for two reasons. First, in these bands
\Beq
\left[\Re(\mu_k)/H\right]^1\propto \mpl/|\bar{\phi}|\qquad |\bar{\phi}|\ll M\,.
\Eeq
Furthermore, the modes tend to spend a lot of time in these narrow bands. This effect can be understood by considering the white flow lines in Fig. \ref{fig:Floq}. The flow lines cross the first narrow band from right to left ($n<2$), left to right ($n>2$), or never leave it ($n=2$). The narrow resonance will clearly persist until non-linear effects become important in the $n=2$ case. Upon closer inspection, the same holds for the $n<2$ and  $n>2$ cases as well. For these two cases, $|\dot{\kappa}|\sim H\kappa$. Since $H$ is decreasing, at some point a given $k$-mode will spend sufficient time within the narrow band for fluctuations to grow substantially. This eventually leads to backreaction on the condensate and complete fragmentation. The above statements are quite general; however, $n=1$ is special. In this case, the higher order bands become too narrow to allow for significant particle production at late times, thus arresting further fragmentation.\\ \\
\noindent{\em Lattice simulations} --- The presence of linear instabilities eventually leads to significant nonlinear dynamics of the fields. To study these non-linear dynamics we solve the equations of motion $\Box\phi+\partial_{\phi}V=0$ and the Friedmann equation numerically using  a parallelized version of \textit{LatticeEasy} \cite{Felder:2000hq}. We initialize the simulations around the end of inflation with a homogeneous condensate + vacuum fluctuations and evolve them for a few$-10$ $e$-folds of expansion after this instant. We ran different simulations (depending on parameters) with $N=128^3, 256^3, 512^3$, and/or $1024^3$ lattices, with the initial size of the simulation volumes $L \sim ({\rm few}-0.1)H_{\rm inf}^{-1}$. We always terminated the simulations before resolution effects became important. Conservatively, the lattice simulation results should be trusted for the number of e-folds shown in Fig. \ref{fig:EoS}. We also verified that our results are independent of the initial power spectra of field fluctuations on scales which are not resonantly excited during the linear stage. The details of the numerical checks and the evolution of the power spectra will be presented elsewhere.
\\ \\ 
\noindent{\em The Equation of State} --- We now turn our attention to the equation of state parameter defined as \\
\Beq
\label{eq:EoS}
w\equiv\frac{\langle p\rangle_{\rm s} }{\langle \rho\rangle_{\rm s}}=\frac{\langle\dot{\phi}^2/2-(\nabla\phi)^2/6a^2-V\rangle_{\rm s}}{\langle \dot{\phi}^2/2+(\nabla\phi)^2/2a^2+V\rangle_{\rm s} }\,.
\Eeq
Here, $p$ and $\rho$ are the energy density and pressure of the inflaton field respectively. The symbol $\langle \hdots\rangle_{\rm s}$ stands for spatial average. The equation of state is often rapidly oscillating compared to the expansion time scales; a time average over many oscillations should be assumed when we refer to $w$ unless otherwise stated. Note that if the spatially and temporally averaged gradient and kinetic energy densities are equal to each other and dominate over the potential energy density, we get $w=1/3$. 

We find the following results for the equation of state at sufficiently late times:
\Beq
w\rightarrow \begin{cases}
                     0\,  &\quad\text{if } n = 1\,,\\
                     1/3\, &    \quad\text{if } n>1\,,
                   \end{cases}\\
\Eeq
and independent of $M\lesssim \mpl $. We explain the independence from $M$, the special nature of $n = 1$, and the generic behavior for $n>1$ below.

For efficient initial resonance ($M\lesssim 2.5\times 10^{-2}\mpl$) the linear fluctuations grow rapidly and backreact on the condensate.  For $n=1$, meta-stable pseudo solitons (oscillons, see for e.g. \cite{Copeland:1995fq,Amin:2011hj}) are copiously produced within $1$ $e$-fold of expansion. They behave as pressureless dust, $w=0$, and can lead to a long period of matter dominated expansion. See the leftmost panel in Fig. \ref{fig:EoS}. For the $n>1$ case, we still form highly overdense field configurations that dominate the energy density, but they are transients, lasting for about an $e$-fold of expansion. Shortly after the transients decay, the inflaton is completely fragmented with almost no energy remaining in the homogeneous condensate. The field configuration now evolves freely in a turbulent manner (as discussed for $n=2$ in \cite{Micha:2002ey}). Numerically, we find that the kinetic and gradient energies are approximately equal to each other and much greater than the potential energy, implying $w\rightarrow 1/3$ (cf. Fig. \ref{fig:EoS}), and that the field is virialized in the sense that $\langle \dot{\phi}^2/2\rangle_{\textrm{s,t}}=\langle (\nabla\phi)^2/2a^2\rangle_{\textrm{s,t}}+n\langle V\rangle_{\textrm{s,t}}$ holds. We can then get an estimate of the deviation of $w$ from $1/3$: $w-1/3\rightarrow (2/3)(n-2)\times$ the fraction of energy density in the potential energy.
\begin{figure}[t]
	\includegraphics[width=0.42\textwidth]{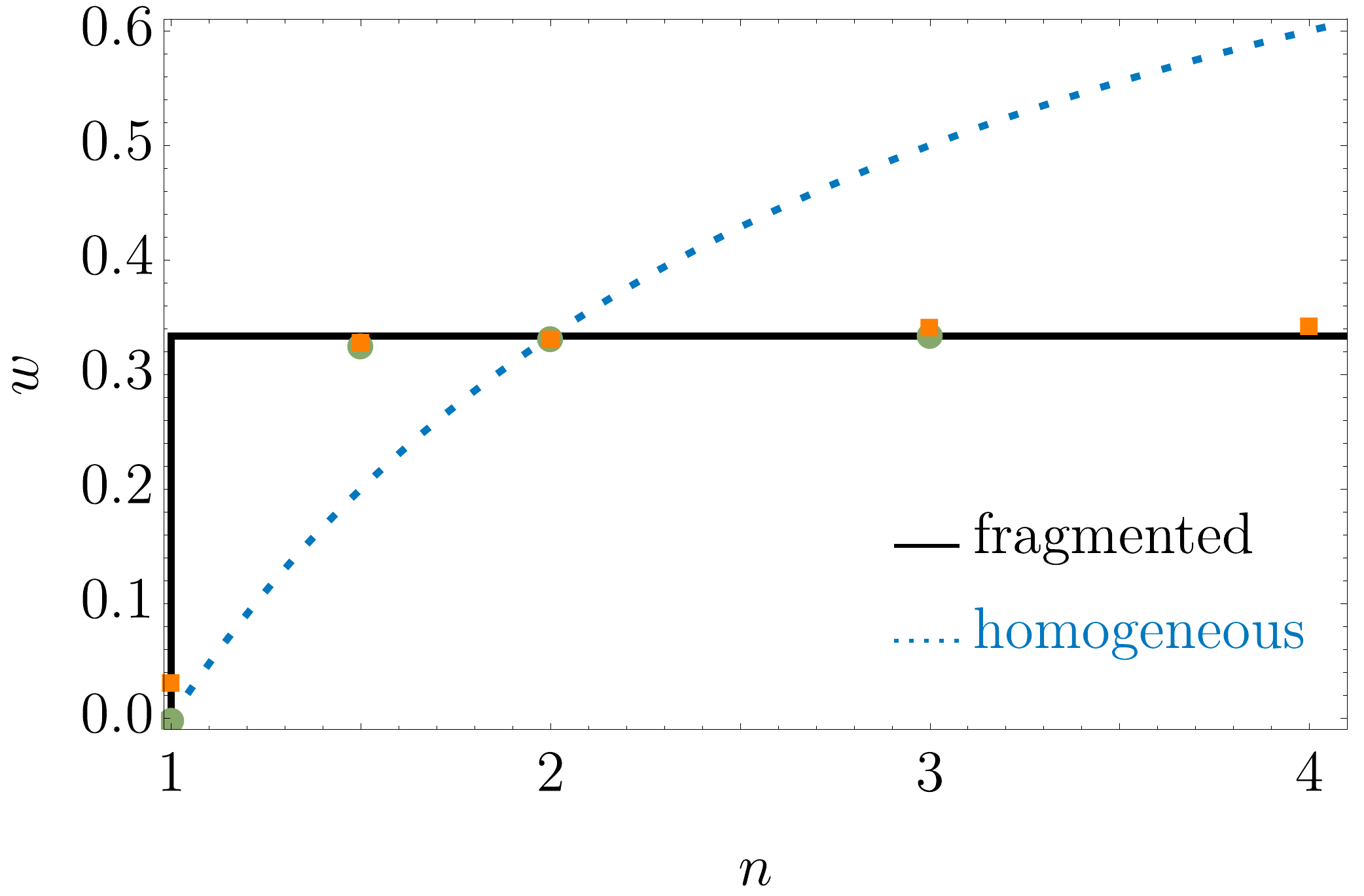}
	\caption{A summary for the {\em asymptotic} equation of state without coupling to additional fields. The numerical results from lattice simulations are shown as green circles for $M\approx 2.45\,\mpl$, and orange squares for $M\approx 7.75\times 10^{-3}\mpl$. The dotted blue line is the expectation from a homogeneous, oscillating condensate.}
	\label{fig:wn}
\end{figure}
For inefficient initial resonance $M\gtrsim 2.5\times 10^{-2}\mpl$ and $n=1$, we observe initially some small excitations of the modes near $k=0$ due to the broad band which is eventually shut off by expansion. The condensate energy is redshifted as $a^{-3}$, slower than the gradient energy ($a^{-4}$). Hence, the fluctuations become ever smaller, and the oscillating condensate determines the equation of state, yielding $w=0$. For $n\!>\!1$, after initial particle production is shut off the condensate energy decays as $a^{-6n/(n+1)}$, whereas the gradient energy stored in field fluctuations decays as $a^{-4}$ (i.e. like radiation) until the first narrow resonance band becomes important and particles are again produced. This second phase of particle production in a narrow $k$ band is expected from our Floquet analysis and confirmed by our lattice simulations. Subsequent evolution includes a shifting of this peak towards higher ($n<2$) or lower ($n>2$) co-moving momenta as expected from the flow lines in the Floquet analysis. This is followed by the generation of a series of secondary peaks from nonlinear scattering (for $n=2$, see \cite{Khlebnikov:1996mc}). Eventually the growth is shut off by backreaction. All the peaks smear out, whereas the remnant condensate continues to oscillate with slowly decaying amplitude, continuing its particle production. After sufficiently long times, we find that the kinetic and gradient energies are approximately equal and much greater than the potential energy with the field again virialized. This yields an equation of state parameter $w \approx1/3$. Note that the $n=2$ case would yield $w=1/3$ for the homogeneous and inhomogeneous field. A summary of the asymptotic equation of state is shown in Fig. \ref{fig:wn}.
\\ \\
\noindent {\em $e$-folds to Radiation Domination} ---
Our linear analysis of the instabilities allows us to estimate the number of $e$-folds after inflation required to reach radiation domination, $\Delta N_{\rm rad}\equiv \int_{a_{\rm end}}^{a_{\rm rad}} d\ln a$, by calculating the time of backreaction of the fluctuations.  First, note that for $n=2$, $\Delta N_{\rm rad}\ll 1$ since in this case $w\rightarrow 1/3$ with and without fragmentation. For all other $n\gtrsim 1$, the universe becomes radiation dominated within 
\Beq
\Delta N_{\rm rad}\sim 
\begin{cases}
1 & M\lesssim 10^{-2}\mpl\,,\\
\dfrac{n+1}{3}\ln \left(\dfrac{\kappa}{\Delta\kappa}\dfrac{10M}{m_{\rm {Pl}}}\right)& M\gtrsim 10^{-2}\mpl\,.\\
\end{cases}
\Eeq
\begin{figure}[t]
	\includegraphics[width=0.45\textwidth]{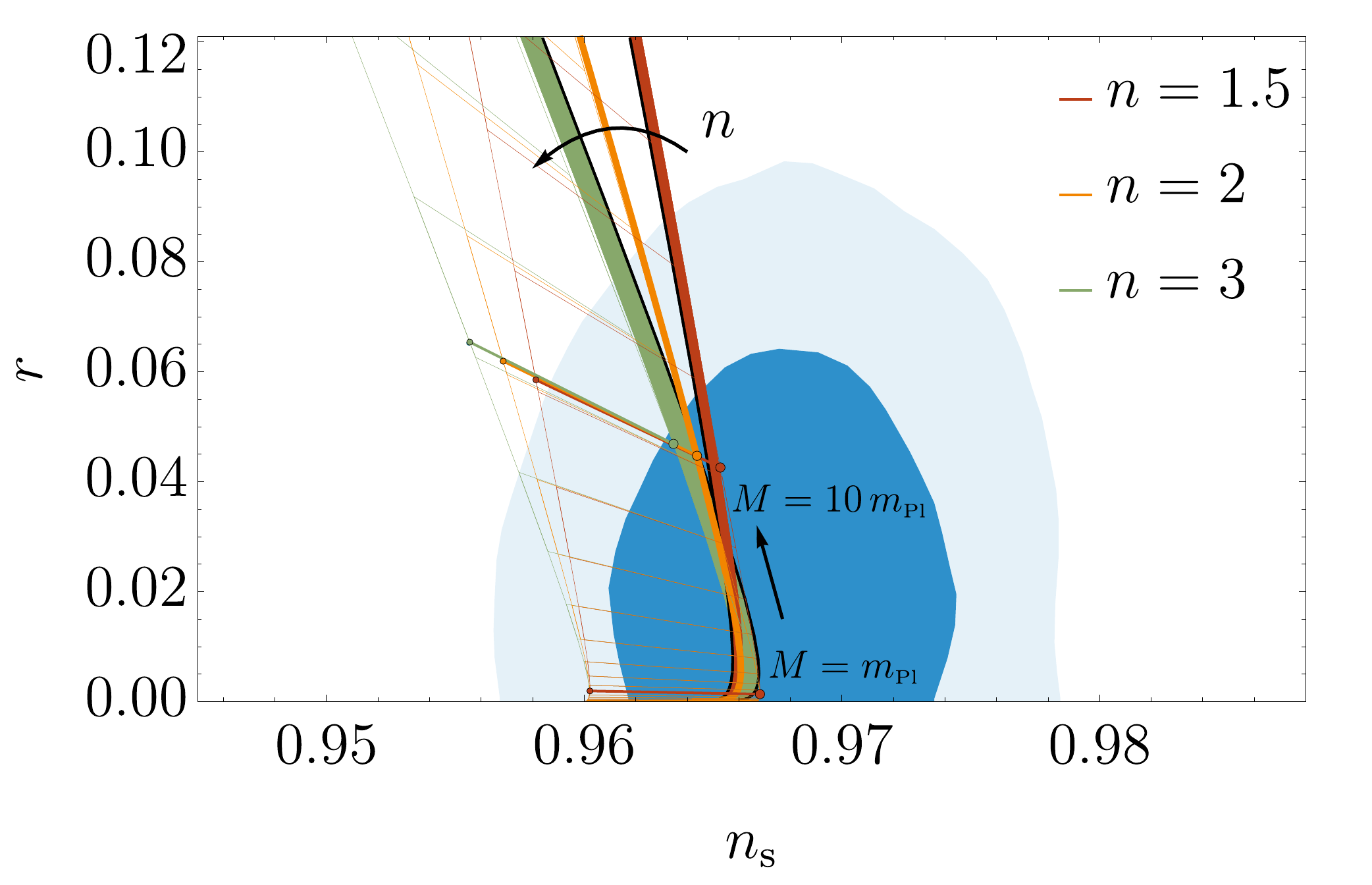}
	\caption{Based on our results, the bounds on $\Delta N_{\rm rad}$ are translated to predictions for $r$ and $n_s$ (filled in colored bands, black edge is for the upper bound). The narrow width of the filled in bands corresponds to a change in $\Delta N_{\rm rad}$ from coupling to other light fields. For comparison, the range of $N_\star=50-60$ commonly used to account for reheating related uncertainties is also shown (thin colored lines, and the ``dumbbells"); the reduction in uncertainty due to our results is significant. Note that $M\gtrsim \mpl$ for most of the above plot. For $M\ll \mpl$ we have $\Delta N_{\rm rad}\lesssim 1$, and $r\ll 10^{-3}$ (and hence, difficult to see in the above observational constraints \cite{Ade:2015lrj}). For the above plot we have focussed on the $\alpha$-attractor models \cite{Kallosh:2013xya,Carrasco:2015pla,Kallosh:2016gqp}, but it can be easily generalized to models with different potentials during the inflationary phase.}
	\label{fig:Planck}
\end{figure}
Here, $\Delta \kappa/\kappa\sim 10^{-2}$ is the fractional width of the first $k\ne 0$ narrow resonance band (cf. Fig. \ref{fig:Floq}). Note that $\Delta \kappa/\kappa$  becomes vanishingly small as $n\rightarrow 1$ (and $n\gg 2$), leading to $\Delta N_{\rm rad}\gg 1$.  These estimates are confirmed by our lattice simulations (see Fig. \ref{fig:EoS}). 

We emphasize that $w\rightarrow1/3$ can be achieved without coupling to other fields for all $n\gtrsim 1$. When coupling to other massless fields is included, $\Delta N_{\rm rad}$ is reduced further. Thus the above calculated $\Delta N_{\rm rad}$ should be taken as an upper bound on $\Delta N_{\rm rad}$. Using these results, we can calculate the expected values of the tensor-to-scalar ratio $r$ and the spectral index $n_{\rm s}$ for different values of $M$ and $n$, even including the uncertainty from couplings to additional light fields (see Fig. \ref{fig:Planck}, we use a pivot scale $k_\star=0.002\,\textrm{Mpc}^{-1}$). The solid black lines use $\Delta N_{\rm rad}$ calculated above, whereas the width of the filled bands allows for a faster approach to radiation domination due to couplings to other fields. We assume the effective bosonic degrees of freedom in the universe at the moment when it reached thermal equilibrium $g_{\rm th}\approx 10^3$; however, changing $g_{\rm th}$ within reasonable bounds does not introduce significant uncertainties. For $n>1$, our conclusions should still hold even if the inflaton has an additional small mass as long as this mass is much smaller than the effective mass due to the curvature of the potential during the approach to radiation domination. Eventually, the small mass might play a role in later time decays.

Note that the $n=1$ case is special and is not shown in the $r-n_s$ plot. When coupling to other massless fields is included, the dynamics can be quite complex, especially for $M\ll \mpl$ due to the existence of oscillons \cite{Amin:2011hj,Gleiser:2011xj,Adshead:2015pva}. For general $n$, the inclusion of additional decay channels and non-minimal couplings \cite{GarciaBellido:2008ab,DeCross:2015uza,Repond:2016sol}, gravitational effects \cite{Khlopov:1985jw,Easther:2010mr} as well as certain quantum aspects \cite{Berges:2015kfa} not captured by our classical simulations can influence predictions from this epoch. 

For the broad class of observationally consistent models considered in this {\em Letter}, our results for the equation of state and the bounds on $\Delta N_{\rm rad}$ are a step towards reducing the model dependence in the reheating epoch and the uncertainty in inflationary and post-inflationary observables.
\\ \\
\noindent {\em Acknowledgments} --- The simulations were performed on the COSMOS Shared Memory system at DAMTP, operated by U. of Cambridge on behalf of the STFC DiRAC HPC Facility. We thank D. Sijacki for her generosity regarding the use of her computational resources under the Cambridge COSMOS Consortium. We acknowledge and thank A. Linde for a detailed and helpful correspondence regarding the models and their implications, R. Easther for suggesting we include additional information regarding the reduction in theoretical uncertainties, S. Carleston for a careful proof-reading and M. Garcia for a discussion regarding the number of reheating $e$-folds, all of which contributed towards an improved manuscript. We also acknowledge useful discussions with D. Kaiser and M. Drewes regarding non-minimal couplings and perturbative decays respectively. 
\bibliographystyle{apsrev4-1}
\bibliography{bibTheEquationOfStateAfterInflation}

\begin{thebibliography}{38}%
\makeatletter
\providecommand \@ifxundefined [1]{%
 \@ifx{#1\undefined}
}%
\providecommand \@ifnum [1]{%
 \ifnum #1\expandafter \@firstoftwo
 \else \expandafter \@secondoftwo
 \fi
}%
\providecommand \@ifx [1]{%
 \ifx #1\expandafter \@firstoftwo
 \else \expandafter \@secondoftwo
 \fi
}%
\providecommand \natexlab [1]{#1}%
\providecommand \enquote  [1]{``#1''}%
\providecommand \bibnamefont  [1]{#1}%
\providecommand \bibfnamefont [1]{#1}%
\providecommand \citenamefont [1]{#1}%
\providecommand \href@noop [0]{\@secondoftwo}%
\providecommand \href [0]{\begingroup \@sanitize@url \@href}%
\providecommand \@href[1]{\@@startlink{#1}\@@href}%
\providecommand \@@href[1]{\endgroup#1\@@endlink}%
\providecommand \@sanitize@url [0]{\catcode `\\12\catcode `\$12\catcode
  `\&12\catcode `\#12\catcode `\^12\catcode `\_12\catcode `\%12\relax}%
\providecommand \@@startlink[1]{}%
\providecommand \@@endlink[0]{}%
\providecommand \url  [0]{\begingroup\@sanitize@url \@url }%
\providecommand \@url [1]{\endgroup\@href {#1}{\urlprefix }}%
\providecommand \urlprefix  [0]{URL }%
\providecommand \Eprint [0]{\href }%
\providecommand \doibase [0]{http://dx.doi.org/}%
\providecommand \selectlanguage [0]{\@gobble}%
\providecommand \bibinfo  [0]{\@secondoftwo}%
\providecommand \bibfield  [0]{\@secondoftwo}%
\providecommand \translation [1]{[#1]}%
\providecommand \BibitemOpen [0]{}%
\providecommand \bibitemStop [0]{}%
\providecommand \bibitemNoStop [0]{.\EOS\space}%
\providecommand \EOS [0]{\spacefactor3000\relax}%
\providecommand \BibitemShut  [1]{\csname bibitem#1\endcsname}%
\let\auto@bib@innerbib\@empty
\bibitem [{\citenamefont {Ade}\ \emph {et~al.}(2015)\citenamefont {Ade} \emph
  {et~al.}}]{Ade:2015lrj}%
  \BibitemOpen
  \bibfield  {author} {\bibinfo {author} {\bibfnamefont {P.~A.~R.}\
  \bibnamefont {Ade}} \emph {et~al.} (\bibinfo {collaboration} {Planck}),\
  }\href@noop {} {\  (\bibinfo {year} {2015})},\ \Eprint
  {http://arxiv.org/abs/1502.02114} {arXiv:1502.02114 [astro-ph.CO]}
  \BibitemShut {NoStop}%
\bibitem [{\citenamefont {Liddle}\ and\ \citenamefont
  {Leach}(2003)}]{Liddle:2003as}%
  \BibitemOpen
  \bibfield  {author} {\bibinfo {author} {\bibfnamefont {A.~R.}\ \bibnamefont
  {Liddle}}\ and\ \bibinfo {author} {\bibfnamefont {S.~M.}\ \bibnamefont
  {Leach}},\ }\href {\doibase 10.1103/PhysRevD.68.103503} {\bibfield  {journal}
  {\bibinfo  {journal} {Phys. Rev.}\ }\textbf {\bibinfo {volume} {D68}},\
  \bibinfo {pages} {103503} (\bibinfo {year} {2003})},\ \Eprint
  {http://arxiv.org/abs/astro-ph/0305263} {arXiv:astro-ph/0305263 [astro-ph]}
  \BibitemShut {NoStop}%
\bibitem [{\citenamefont {Adshead}\ \emph {et~al.}(2011)\citenamefont
  {Adshead}, \citenamefont {Easther}, \citenamefont {Pritchard},\ and\
  \citenamefont {Loeb}}]{Adshead:2010mc}%
  \BibitemOpen
  \bibfield  {author} {\bibinfo {author} {\bibfnamefont {P.}~\bibnamefont
  {Adshead}}, \bibinfo {author} {\bibfnamefont {R.}~\bibnamefont {Easther}},
  \bibinfo {author} {\bibfnamefont {J.}~\bibnamefont {Pritchard}}, \ and\
  \bibinfo {author} {\bibfnamefont {A.}~\bibnamefont {Loeb}},\ }\href {\doibase
  10.1088/1475-7516/2011/02/021} {\bibfield  {journal} {\bibinfo  {journal}
  {JCAP}\ }\textbf {\bibinfo {volume} {1102}},\ \bibinfo {pages} {021}
  (\bibinfo {year} {2011})},\ \Eprint {http://arxiv.org/abs/1007.3748}
  {arXiv:1007.3748 [astro-ph.CO]} \BibitemShut {NoStop}%
\bibitem [{\citenamefont {Creminelli}\ \emph {et~al.}(2014)\citenamefont
  {Creminelli}, \citenamefont {López~Nacir}, \citenamefont {Simonović},
  \citenamefont {Trevisan},\ and\ \citenamefont
  {Zaldarriaga}}]{Creminelli:2014oaa}%
  \BibitemOpen
  \bibfield  {author} {\bibinfo {author} {\bibfnamefont {P.}~\bibnamefont
  {Creminelli}}, \bibinfo {author} {\bibfnamefont {D.}~\bibnamefont
  {López~Nacir}}, \bibinfo {author} {\bibfnamefont {M.}~\bibnamefont
  {Simonović}}, \bibinfo {author} {\bibfnamefont {G.}~\bibnamefont
  {Trevisan}}, \ and\ \bibinfo {author} {\bibfnamefont {M.}~\bibnamefont
  {Zaldarriaga}},\ }\href {\doibase 10.1103/PhysRevLett.112.241303} {\bibfield
  {journal} {\bibinfo  {journal} {Phys. Rev. Lett.}\ }\textbf {\bibinfo
  {volume} {112}},\ \bibinfo {pages} {241303} (\bibinfo {year} {2014})},\
  \Eprint {http://arxiv.org/abs/1404.1065} {arXiv:1404.1065 [astro-ph.CO]}
  \BibitemShut {NoStop}%
\bibitem [{\citenamefont {Dai}\ \emph {et~al.}(2014)\citenamefont {Dai},
  \citenamefont {Kamionkowski},\ and\ \citenamefont {Wang}}]{Dai:2014jja}%
  \BibitemOpen
  \bibfield  {author} {\bibinfo {author} {\bibfnamefont {L.}~\bibnamefont
  {Dai}}, \bibinfo {author} {\bibfnamefont {M.}~\bibnamefont {Kamionkowski}}, \
  and\ \bibinfo {author} {\bibfnamefont {J.}~\bibnamefont {Wang}},\ }\href
  {\doibase 10.1103/PhysRevLett.113.041302} {\bibfield  {journal} {\bibinfo
  {journal} {Phys. Rev. Lett.}\ }\textbf {\bibinfo {volume} {113}},\ \bibinfo
  {pages} {041302} (\bibinfo {year} {2014})},\ \Eprint
  {http://arxiv.org/abs/1404.6704} {arXiv:1404.6704 [astro-ph.CO]} \BibitemShut
  {NoStop}%
\bibitem [{\citenamefont {Martin}\ \emph {et~al.}(2015)\citenamefont {Martin},
  \citenamefont {Ringeval},\ and\ \citenamefont {Vennin}}]{Martin:2014nya}%
  \BibitemOpen
  \bibfield  {author} {\bibinfo {author} {\bibfnamefont {J.}~\bibnamefont
  {Martin}}, \bibinfo {author} {\bibfnamefont {C.}~\bibnamefont {Ringeval}}, \
  and\ \bibinfo {author} {\bibfnamefont {V.}~\bibnamefont {Vennin}},\ }\href
  {\doibase 10.1103/PhysRevLett.114.081303} {\bibfield  {journal} {\bibinfo
  {journal} {Phys. Rev. Lett.}\ }\textbf {\bibinfo {volume} {114}},\ \bibinfo
  {pages} {081303} (\bibinfo {year} {2015})},\ \Eprint
  {http://arxiv.org/abs/1410.7958} {arXiv:1410.7958 [astro-ph.CO]} \BibitemShut
  {NoStop}%
\bibitem [{\citenamefont {Munoz}\ and\ \citenamefont
  {Kamionkowski}(2015)}]{Munoz:2014eqa}%
  \BibitemOpen
  \bibfield  {author} {\bibinfo {author} {\bibfnamefont {J.~B.}\ \bibnamefont
  {Munoz}}\ and\ \bibinfo {author} {\bibfnamefont {M.}~\bibnamefont
  {Kamionkowski}},\ }\href {\doibase 10.1103/PhysRevD.91.043521} {\bibfield
  {journal} {\bibinfo  {journal} {Phys. Rev.}\ }\textbf {\bibinfo {volume}
  {D91}},\ \bibinfo {pages} {043521} (\bibinfo {year} {2015})},\ \Eprint
  {http://arxiv.org/abs/1412.0656} {arXiv:1412.0656 [astro-ph.CO]} \BibitemShut
  {NoStop}%
\bibitem [{\citenamefont {Cook}\ \emph {et~al.}(2015)\citenamefont {Cook},
  \citenamefont {Dimastrogiovanni}, \citenamefont {Easson},\ and\ \citenamefont
  {Krauss}}]{Cook:2015vqa}%
  \BibitemOpen
  \bibfield  {author} {\bibinfo {author} {\bibfnamefont {J.~L.}\ \bibnamefont
  {Cook}}, \bibinfo {author} {\bibfnamefont {E.}~\bibnamefont
  {Dimastrogiovanni}}, \bibinfo {author} {\bibfnamefont {D.~A.}\ \bibnamefont
  {Easson}}, \ and\ \bibinfo {author} {\bibfnamefont {L.~M.}\ \bibnamefont
  {Krauss}},\ }\href {\doibase 10.1088/1475-7516/2015/04/047} {\bibfield
  {journal} {\bibinfo  {journal} {JCAP}\ }\textbf {\bibinfo {volume} {1504}},\
  \bibinfo {pages} {047} (\bibinfo {year} {2015})},\ \Eprint
  {http://arxiv.org/abs/1502.04673} {arXiv:1502.04673 [astro-ph.CO]}
  \BibitemShut {NoStop}%
\bibitem [{\citenamefont {Ellis}\ \emph {et~al.}(2015)\citenamefont {Ellis},
  \citenamefont {Garcia}, \citenamefont {Nanopoulos},\ and\ \citenamefont
  {Olive}}]{Ellis:2015pla}%
  \BibitemOpen
  \bibfield  {author} {\bibinfo {author} {\bibfnamefont {J.}~\bibnamefont
  {Ellis}}, \bibinfo {author} {\bibfnamefont {M.~A.~G.}\ \bibnamefont
  {Garcia}}, \bibinfo {author} {\bibfnamefont {D.~V.}\ \bibnamefont
  {Nanopoulos}}, \ and\ \bibinfo {author} {\bibfnamefont {K.~A.}\ \bibnamefont
  {Olive}},\ }\href {\doibase 10.1088/1475-7516/2015/07/050} {\bibfield
  {journal} {\bibinfo  {journal} {JCAP}\ }\textbf {\bibinfo {volume} {1507}},\
  \bibinfo {pages} {050} (\bibinfo {year} {2015})},\ \Eprint
  {http://arxiv.org/abs/1505.06986} {arXiv:1505.06986 [hep-ph]} \BibitemShut
  {NoStop}%
\bibitem [{\citenamefont {Ueno}\ and\ \citenamefont
  {Yamamoto}(2016)}]{Ueno:2016dim}%
  \BibitemOpen
  \bibfield  {author} {\bibinfo {author} {\bibfnamefont {Y.}~\bibnamefont
  {Ueno}}\ and\ \bibinfo {author} {\bibfnamefont {K.}~\bibnamefont
  {Yamamoto}},\ }\href@noop {} {\  (\bibinfo {year} {2016})},\ \Eprint
  {http://arxiv.org/abs/1602.07427} {arXiv:1602.07427 [astro-ph.CO]}
  \BibitemShut {NoStop}%
\bibitem [{\citenamefont {Eshaghi}\ \emph {et~al.}(2016)\citenamefont
  {Eshaghi}, \citenamefont {Zarei}, \citenamefont {Riazi},\ and\ \citenamefont
  {Kiasatpour}}]{Eshaghi:2016kne}%
  \BibitemOpen
  \bibfield  {author} {\bibinfo {author} {\bibfnamefont {M.}~\bibnamefont
  {Eshaghi}}, \bibinfo {author} {\bibfnamefont {M.}~\bibnamefont {Zarei}},
  \bibinfo {author} {\bibfnamefont {N.}~\bibnamefont {Riazi}}, \ and\ \bibinfo
  {author} {\bibfnamefont {A.}~\bibnamefont {Kiasatpour}},\ }\href@noop {} {\
  (\bibinfo {year} {2016})},\ \Eprint {http://arxiv.org/abs/1602.07914}
  {arXiv:1602.07914 [astro-ph.CO]} \BibitemShut {NoStop}%
\bibitem [{\citenamefont {Giudice}\ \emph {et~al.}(1999)\citenamefont
  {Giudice}, \citenamefont {Tkachev},\ and\ \citenamefont
  {Riotto}}]{Giudice:1999yt}%
  \BibitemOpen
  \bibfield  {author} {\bibinfo {author} {\bibfnamefont {G.~F.}\ \bibnamefont
  {Giudice}}, \bibinfo {author} {\bibfnamefont {I.}~\bibnamefont {Tkachev}}, \
  and\ \bibinfo {author} {\bibfnamefont {A.}~\bibnamefont {Riotto}},\ }\href
  {\doibase 10.1088/1126-6708/1999/08/009} {\bibfield  {journal} {\bibinfo
  {journal} {JHEP}\ }\textbf {\bibinfo {volume} {08}},\ \bibinfo {pages} {009}
  (\bibinfo {year} {1999})},\ \Eprint {http://arxiv.org/abs/hep-ph/9907510}
  {arXiv:hep-ph/9907510 [hep-ph]} \BibitemShut {NoStop}%
\bibitem [{\citenamefont {Hertzberg}\ and\ \citenamefont
  {Karouby}(2014)}]{Hertzberg:2013jba}%
  \BibitemOpen
  \bibfield  {author} {\bibinfo {author} {\bibfnamefont {M.~P.}\ \bibnamefont
  {Hertzberg}}\ and\ \bibinfo {author} {\bibfnamefont {J.}~\bibnamefont
  {Karouby}},\ }\href {\doibase 10.1016/j.physletb.2014.08.021} {\bibfield
  {journal} {\bibinfo  {journal} {Phys. Lett.}\ }\textbf {\bibinfo {volume}
  {B737}},\ \bibinfo {pages} {34} (\bibinfo {year} {2014})},\ \Eprint
  {http://arxiv.org/abs/1309.0007} {arXiv:1309.0007 [hep-ph]} \BibitemShut
  {NoStop}%
\bibitem [{\citenamefont {Kane}\ \emph {et~al.}(2015)\citenamefont {Kane},
  \citenamefont {Sinha},\ and\ \citenamefont {Watson}}]{Kane:2015jia}%
  \BibitemOpen
  \bibfield  {author} {\bibinfo {author} {\bibfnamefont {G.}~\bibnamefont
  {Kane}}, \bibinfo {author} {\bibfnamefont {K.}~\bibnamefont {Sinha}}, \ and\
  \bibinfo {author} {\bibfnamefont {S.}~\bibnamefont {Watson}},\ }\href
  {\doibase 10.1142/S0218271815300220} {\bibfield  {journal} {\bibinfo
  {journal} {Int. J. Mod. Phys.}\ }\textbf {\bibinfo {volume} {D24}},\ \bibinfo
  {pages} {1530022} (\bibinfo {year} {2015})},\ \Eprint
  {http://arxiv.org/abs/1502.07746} {arXiv:1502.07746 [hep-th]} \BibitemShut
  {NoStop}%
\bibitem [{\citenamefont {Turner}(1983)}]{Turner:1983he}%
  \BibitemOpen
  \bibfield  {author} {\bibinfo {author} {\bibfnamefont {M.~S.}\ \bibnamefont
  {Turner}},\ }\href {\doibase 10.1103/PhysRevD.28.1243} {\bibfield  {journal}
  {\bibinfo  {journal} {Phys. Rev.}\ }\textbf {\bibinfo {volume} {D28}},\
  \bibinfo {pages} {1243} (\bibinfo {year} {1983})}\BibitemShut {NoStop}%
\bibitem [{\citenamefont {Podolsky}\ \emph {et~al.}(2006)\citenamefont
  {Podolsky}, \citenamefont {Felder}, \citenamefont {Kofman},\ and\
  \citenamefont {Peloso}}]{Podolsky:2005bw}%
  \BibitemOpen
  \bibfield  {author} {\bibinfo {author} {\bibfnamefont {D.~I.}\ \bibnamefont
  {Podolsky}}, \bibinfo {author} {\bibfnamefont {G.~N.}\ \bibnamefont
  {Felder}}, \bibinfo {author} {\bibfnamefont {L.}~\bibnamefont {Kofman}}, \
  and\ \bibinfo {author} {\bibfnamefont {M.}~\bibnamefont {Peloso}},\ }\href
  {\doibase 10.1103/PhysRevD.73.023501} {\bibfield  {journal} {\bibinfo
  {journal} {Phys. Rev.}\ }\textbf {\bibinfo {volume} {D73}},\ \bibinfo {pages}
  {023501} (\bibinfo {year} {2006})},\ \Eprint
  {http://arxiv.org/abs/hep-ph/0507096} {arXiv:hep-ph/0507096 [hep-ph]}
  \BibitemShut {NoStop}%
\bibitem [{\citenamefont {Kallosh}\ and\ \citenamefont
  {Linde}(2013)}]{Kallosh:2013xya}%
  \BibitemOpen
  \bibfield  {author} {\bibinfo {author} {\bibfnamefont {R.}~\bibnamefont
  {Kallosh}}\ and\ \bibinfo {author} {\bibfnamefont {A.}~\bibnamefont
  {Linde}},\ }\href {\doibase 10.1088/1475-7516/2013/06/028} {\bibfield
  {journal} {\bibinfo  {journal} {JCAP}\ }\textbf {\bibinfo {volume} {1306}},\
  \bibinfo {pages} {028} (\bibinfo {year} {2013})},\ \Eprint
  {http://arxiv.org/abs/1306.3214} {arXiv:1306.3214 [hep-th]} \BibitemShut
  {NoStop}%
\bibitem [{\citenamefont {Carrasco}\ \emph {et~al.}(2015)\citenamefont
  {Carrasco}, \citenamefont {Kallosh},\ and\ \citenamefont
  {Linde}}]{Carrasco:2015pla}%
  \BibitemOpen
  \bibfield  {author} {\bibinfo {author} {\bibfnamefont {J.~J.~M.}\
  \bibnamefont {Carrasco}}, \bibinfo {author} {\bibfnamefont {R.}~\bibnamefont
  {Kallosh}}, \ and\ \bibinfo {author} {\bibfnamefont {A.}~\bibnamefont
  {Linde}},\ }\href {\doibase 10.1007/JHEP10(2015)147} {\bibfield  {journal}
  {\bibinfo  {journal} {JHEP}\ }\textbf {\bibinfo {volume} {10}},\ \bibinfo
  {pages} {147} (\bibinfo {year} {2015})},\ \Eprint
  {http://arxiv.org/abs/1506.01708} {arXiv:1506.01708 [hep-th]} \BibitemShut
  {NoStop}%
\bibitem [{\citenamefont {Kallosh}\ and\ \citenamefont
  {Linde}(2016)}]{Kallosh:2016gqp}%
  \BibitemOpen
  \bibfield  {author} {\bibinfo {author} {\bibfnamefont {R.}~\bibnamefont
  {Kallosh}}\ and\ \bibinfo {author} {\bibfnamefont {A.}~\bibnamefont
  {Linde}},\ }\href@noop {} {\  (\bibinfo {year} {2016})},\ \Eprint
  {http://arxiv.org/abs/1604.00444} {arXiv:1604.00444 [hep-th]} \BibitemShut
  {NoStop}%
\bibitem [{\citenamefont {Silverstein}\ and\ \citenamefont
  {Westphal}(2008)}]{Silverstein:2008sg}%
  \BibitemOpen
  \bibfield  {author} {\bibinfo {author} {\bibfnamefont {E.}~\bibnamefont
  {Silverstein}}\ and\ \bibinfo {author} {\bibfnamefont {A.}~\bibnamefont
  {Westphal}},\ }\href {\doibase 10.1103/PhysRevD.78.106003} {\bibfield
  {journal} {\bibinfo  {journal} {Phys. Rev.}\ }\textbf {\bibinfo {volume}
  {D78}},\ \bibinfo {pages} {106003} (\bibinfo {year} {2008})},\ \Eprint
  {http://arxiv.org/abs/0803.3085} {arXiv:0803.3085 [hep-th]} \BibitemShut
  {NoStop}%
\bibitem [{\citenamefont {McAllister}\ \emph {et~al.}(2014)\citenamefont
  {McAllister}, \citenamefont {Silverstein}, \citenamefont {Westphal},\ and\
  \citenamefont {Wrase}}]{McAllister:2014mpa}%
  \BibitemOpen
  \bibfield  {author} {\bibinfo {author} {\bibfnamefont {L.}~\bibnamefont
  {McAllister}}, \bibinfo {author} {\bibfnamefont {E.}~\bibnamefont
  {Silverstein}}, \bibinfo {author} {\bibfnamefont {A.}~\bibnamefont
  {Westphal}}, \ and\ \bibinfo {author} {\bibfnamefont {T.}~\bibnamefont
  {Wrase}},\ }\href {\doibase 10.1007/JHEP09(2014)123} {\bibfield  {journal}
  {\bibinfo  {journal} {JHEP}\ }\textbf {\bibinfo {volume} {09}},\ \bibinfo
  {pages} {123} (\bibinfo {year} {2014})},\ \Eprint
  {http://arxiv.org/abs/1405.3652} {arXiv:1405.3652 [hep-th]} \BibitemShut
  {NoStop}%
\bibitem [{\citenamefont {Kofman}\ \emph {et~al.}(1994)\citenamefont {Kofman},
  \citenamefont {Linde},\ and\ \citenamefont {Starobinsky}}]{Kofman:1994rk}%
  \BibitemOpen
  \bibfield  {author} {\bibinfo {author} {\bibfnamefont {L.}~\bibnamefont
  {Kofman}}, \bibinfo {author} {\bibfnamefont {A.~D.}\ \bibnamefont {Linde}}, \
  and\ \bibinfo {author} {\bibfnamefont {A.~A.}\ \bibnamefont {Starobinsky}},\
  }\href {\doibase 10.1103/PhysRevLett.73.3195} {\bibfield  {journal} {\bibinfo
   {journal} {Phys. Rev. Lett.}\ }\textbf {\bibinfo {volume} {73}},\ \bibinfo
  {pages} {3195} (\bibinfo {year} {1994})},\ \Eprint
  {http://arxiv.org/abs/hep-th/9405187} {arXiv:hep-th/9405187 [hep-th]}
  \BibitemShut {NoStop}%
\bibitem [{\citenamefont {Shtanov}\ \emph {et~al.}(1995)\citenamefont
  {Shtanov}, \citenamefont {Traschen},\ and\ \citenamefont
  {Brandenberger}}]{Shtanov:1994ce}%
  \BibitemOpen
  \bibfield  {author} {\bibinfo {author} {\bibfnamefont {Y.}~\bibnamefont
  {Shtanov}}, \bibinfo {author} {\bibfnamefont {J.~H.}\ \bibnamefont
  {Traschen}}, \ and\ \bibinfo {author} {\bibfnamefont {R.~H.}\ \bibnamefont
  {Brandenberger}},\ }\href {\doibase 10.1103/PhysRevD.51.5438} {\bibfield
  {journal} {\bibinfo  {journal} {Phys. Rev.}\ }\textbf {\bibinfo {volume}
  {D51}},\ \bibinfo {pages} {5438} (\bibinfo {year} {1995})},\ \Eprint
  {http://arxiv.org/abs/hep-ph/9407247} {arXiv:hep-ph/9407247 [hep-ph]}
  \BibitemShut {NoStop}%
\bibitem [{\citenamefont {Kofman}\ \emph {et~al.}(1997)\citenamefont {Kofman},
  \citenamefont {Linde},\ and\ \citenamefont {Starobinsky}}]{Kofman1997}%
  \BibitemOpen
  \bibfield  {author} {\bibinfo {author} {\bibfnamefont {L.}~\bibnamefont
  {Kofman}}, \bibinfo {author} {\bibfnamefont {A.~D.}\ \bibnamefont {Linde}}, \
  and\ \bibinfo {author} {\bibfnamefont {A.~A.}\ \bibnamefont {Starobinsky}},\
  }\href {\doibase 10.1103/PhysRevD.56.3258} {\bibfield  {journal} {\bibinfo
  {journal} {Phys.Rev.}\ }\textbf {\bibinfo {volume} {D56}},\ \bibinfo {pages}
  {3258} (\bibinfo {year} {1997})},\ \Eprint
  {http://arxiv.org/abs/hep-ph/9704452} {arXiv:hep-ph/9704452 [hep-ph]}
  \BibitemShut {NoStop}%
\bibitem [{\citenamefont {Amin}\ \emph {et~al.}(2014)\citenamefont {Amin},
  \citenamefont {Hertzberg}, \citenamefont {Kaiser},\ and\ \citenamefont
  {Karouby}}]{Amin2014}%
  \BibitemOpen
  \bibfield  {author} {\bibinfo {author} {\bibfnamefont {M.~A.}\ \bibnamefont
  {Amin}}, \bibinfo {author} {\bibfnamefont {M.~P.}\ \bibnamefont {Hertzberg}},
  \bibinfo {author} {\bibfnamefont {D.~I.}\ \bibnamefont {Kaiser}}, \ and\
  \bibinfo {author} {\bibfnamefont {J.}~\bibnamefont {Karouby}},\ }\href
  {\doibase 10.1142/S0218271815300037} {\bibfield  {journal} {\bibinfo
  {journal} {Int.J.Mod.Phys.}\ }\textbf {\bibinfo {volume} {D24}},\ \bibinfo
  {pages} {1530003} (\bibinfo {year} {2014})},\ \Eprint
  {http://arxiv.org/abs/1410.3808} {arXiv:1410.3808 [hep-ph]} \BibitemShut
  {NoStop}%
\bibitem [{\citenamefont {Felder}\ and\ \citenamefont
  {Tkachev}(2008)}]{Felder:2000hq}%
  \BibitemOpen
  \bibfield  {author} {\bibinfo {author} {\bibfnamefont {G.~N.}\ \bibnamefont
  {Felder}}\ and\ \bibinfo {author} {\bibfnamefont {I.}~\bibnamefont
  {Tkachev}},\ }\href {\doibase 10.1016/j.cpc.2008.02.009} {\bibfield
  {journal} {\bibinfo  {journal} {Comput. Phys. Commun.}\ }\textbf {\bibinfo
  {volume} {178}},\ \bibinfo {pages} {929} (\bibinfo {year} {2008})},\ \Eprint
  {http://arxiv.org/abs/hep-ph/0011159} {arXiv:hep-ph/0011159 [hep-ph]}
  \BibitemShut {NoStop}%
\bibitem [{\citenamefont {Copeland}\ \emph {et~al.}(1995)\citenamefont
  {Copeland}, \citenamefont {Gleiser},\ and\ \citenamefont
  {Muller}}]{Copeland:1995fq}%
  \BibitemOpen
  \bibfield  {author} {\bibinfo {author} {\bibfnamefont {E.~J.}\ \bibnamefont
  {Copeland}}, \bibinfo {author} {\bibfnamefont {M.}~\bibnamefont {Gleiser}}, \
  and\ \bibinfo {author} {\bibfnamefont {H.~R.}\ \bibnamefont {Muller}},\
  }\href {\doibase 10.1103/PhysRevD.52.1920} {\bibfield  {journal} {\bibinfo
  {journal} {Phys. Rev.}\ }\textbf {\bibinfo {volume} {D52}},\ \bibinfo {pages}
  {1920} (\bibinfo {year} {1995})},\ \Eprint
  {http://arxiv.org/abs/hep-ph/9503217} {arXiv:hep-ph/9503217 [hep-ph]}
  \BibitemShut {NoStop}%
\bibitem [{\citenamefont {Amin}\ \emph {et~al.}(2012)\citenamefont {Amin},
  \citenamefont {Easther}, \citenamefont {Finkel}, \citenamefont {Flauger},\
  and\ \citenamefont {Hertzberg}}]{Amin:2011hj}%
  \BibitemOpen
  \bibfield  {author} {\bibinfo {author} {\bibfnamefont {M.~A.}\ \bibnamefont
  {Amin}}, \bibinfo {author} {\bibfnamefont {R.}~\bibnamefont {Easther}},
  \bibinfo {author} {\bibfnamefont {H.}~\bibnamefont {Finkel}}, \bibinfo
  {author} {\bibfnamefont {R.}~\bibnamefont {Flauger}}, \ and\ \bibinfo
  {author} {\bibfnamefont {M.~P.}\ \bibnamefont {Hertzberg}},\ }\href {\doibase
  10.1103/PhysRevLett.108.241302} {\bibfield  {journal} {\bibinfo  {journal}
  {Phys. Rev. Lett.}\ }\textbf {\bibinfo {volume} {108}},\ \bibinfo {pages}
  {241302} (\bibinfo {year} {2012})},\ \Eprint {http://arxiv.org/abs/1106.3335}
  {arXiv:1106.3335 [astro-ph.CO]} \BibitemShut {NoStop}%
\bibitem [{\citenamefont {Micha}\ and\ \citenamefont
  {Tkachev}(2003)}]{Micha:2002ey}%
  \BibitemOpen
  \bibfield  {author} {\bibinfo {author} {\bibfnamefont {R.}~\bibnamefont
  {Micha}}\ and\ \bibinfo {author} {\bibfnamefont {I.~I.}\ \bibnamefont
  {Tkachev}},\ }\href {\doibase 10.1103/PhysRevLett.90.121301} {\bibfield
  {journal} {\bibinfo  {journal} {Phys. Rev. Lett.}\ }\textbf {\bibinfo
  {volume} {90}},\ \bibinfo {pages} {121301} (\bibinfo {year} {2003})},\
  \Eprint {http://arxiv.org/abs/hep-ph/0210202} {arXiv:hep-ph/0210202 [hep-ph]}
  \BibitemShut {NoStop}%
\bibitem [{\citenamefont {Khlebnikov}\ and\ \citenamefont
  {Tkachev}(1996)}]{Khlebnikov:1996mc}%
  \BibitemOpen
  \bibfield  {author} {\bibinfo {author} {\bibfnamefont {S.~{\relax Yu}.}\
  \bibnamefont {Khlebnikov}}\ and\ \bibinfo {author} {\bibfnamefont {I.~I.}\
  \bibnamefont {Tkachev}},\ }\href {\doibase 10.1103/PhysRevLett.77.219}
  {\bibfield  {journal} {\bibinfo  {journal} {Phys. Rev. Lett.}\ }\textbf
  {\bibinfo {volume} {77}},\ \bibinfo {pages} {219} (\bibinfo {year} {1996})},\
  \Eprint {http://arxiv.org/abs/hep-ph/9603378} {arXiv:hep-ph/9603378 [hep-ph]}
  \BibitemShut {NoStop}%
\bibitem [{\citenamefont {Gleiser}\ \emph {et~al.}(2011)\citenamefont
  {Gleiser}, \citenamefont {Graham},\ and\ \citenamefont
  {Stamatopoulos}}]{Gleiser:2011xj}%
  \BibitemOpen
  \bibfield  {author} {\bibinfo {author} {\bibfnamefont {M.}~\bibnamefont
  {Gleiser}}, \bibinfo {author} {\bibfnamefont {N.}~\bibnamefont {Graham}}, \
  and\ \bibinfo {author} {\bibfnamefont {N.}~\bibnamefont {Stamatopoulos}},\
  }\href {\doibase 10.1103/PhysRevD.83.096010} {\bibfield  {journal} {\bibinfo
  {journal} {Phys. Rev.}\ }\textbf {\bibinfo {volume} {D83}},\ \bibinfo {pages}
  {096010} (\bibinfo {year} {2011})},\ \Eprint {http://arxiv.org/abs/1103.1911}
  {arXiv:1103.1911 [hep-th]} \BibitemShut {NoStop}%
\bibitem [{\citenamefont {Adshead}\ \emph {et~al.}(2015)\citenamefont
  {Adshead}, \citenamefont {Giblin}, \citenamefont {Scully},\ and\
  \citenamefont {Sfakianakis}}]{Adshead:2015pva}%
  \BibitemOpen
  \bibfield  {author} {\bibinfo {author} {\bibfnamefont {P.}~\bibnamefont
  {Adshead}}, \bibinfo {author} {\bibfnamefont {J.~T.}\ \bibnamefont {Giblin}},
  \bibinfo {author} {\bibfnamefont {T.~R.}\ \bibnamefont {Scully}}, \ and\
  \bibinfo {author} {\bibfnamefont {E.~I.}\ \bibnamefont {Sfakianakis}},\
  }\href {\doibase 10.1088/1475-7516/2015/12/034} {\bibfield  {journal}
  {\bibinfo  {journal} {JCAP}\ }\textbf {\bibinfo {volume} {1512}},\ \bibinfo
  {pages} {034} (\bibinfo {year} {2015})},\ \Eprint
  {http://arxiv.org/abs/1502.06506} {arXiv:1502.06506 [astro-ph.CO]}
  \BibitemShut {NoStop}%
\bibitem [{\citenamefont {Garcia-Bellido}\ \emph {et~al.}(2009)\citenamefont
  {Garcia-Bellido}, \citenamefont {Figueroa},\ and\ \citenamefont
  {Rubio}}]{GarciaBellido:2008ab}%
  \BibitemOpen
  \bibfield  {author} {\bibinfo {author} {\bibfnamefont {J.}~\bibnamefont
  {Garcia-Bellido}}, \bibinfo {author} {\bibfnamefont {D.~G.}\ \bibnamefont
  {Figueroa}}, \ and\ \bibinfo {author} {\bibfnamefont {J.}~\bibnamefont
  {Rubio}},\ }\href {\doibase 10.1103/PhysRevD.79.063531} {\bibfield  {journal}
  {\bibinfo  {journal} {Phys. Rev.}\ }\textbf {\bibinfo {volume} {D79}},\
  \bibinfo {pages} {063531} (\bibinfo {year} {2009})},\ \Eprint
  {http://arxiv.org/abs/0812.4624} {arXiv:0812.4624 [hep-ph]} \BibitemShut
  {NoStop}%
\bibitem [{\citenamefont {DeCross}\ \emph {et~al.}(2015)\citenamefont
  {DeCross}, \citenamefont {Kaiser}, \citenamefont {Prabhu}, \citenamefont
  {Prescod-Weinstein},\ and\ \citenamefont {Sfakianakis}}]{DeCross:2015uza}%
  \BibitemOpen
  \bibfield  {author} {\bibinfo {author} {\bibfnamefont {M.~P.}\ \bibnamefont
  {DeCross}}, \bibinfo {author} {\bibfnamefont {D.~I.}\ \bibnamefont {Kaiser}},
  \bibinfo {author} {\bibfnamefont {A.}~\bibnamefont {Prabhu}}, \bibinfo
  {author} {\bibfnamefont {C.}~\bibnamefont {Prescod-Weinstein}}, \ and\
  \bibinfo {author} {\bibfnamefont {E.~I.}\ \bibnamefont {Sfakianakis}},\
  }\href@noop {} {\  (\bibinfo {year} {2015})},\ \Eprint
  {http://arxiv.org/abs/1510.08553} {arXiv:1510.08553 [astro-ph.CO]}
  \BibitemShut {NoStop}%
\bibitem [{\citenamefont {Repond}\ and\ \citenamefont
  {Rubio}(2016)}]{Repond:2016sol}%
  \BibitemOpen
  \bibfield  {author} {\bibinfo {author} {\bibfnamefont {J.}~\bibnamefont
  {Repond}}\ and\ \bibinfo {author} {\bibfnamefont {J.}~\bibnamefont {Rubio}},\
  }\href {\doibase 10.1088/1475-7516/2016/07/043} {\bibfield  {journal}
  {\bibinfo  {journal} {JCAP}\ }\textbf {\bibinfo {volume} {1607}},\ \bibinfo
  {pages} {043} (\bibinfo {year} {2016})},\ \Eprint
  {http://arxiv.org/abs/1604.08238} {arXiv:1604.08238 [astro-ph.CO]}
  \BibitemShut {NoStop}%
\bibitem [{\citenamefont {Khlopov}\ \emph {et~al.}(1985)\citenamefont
  {Khlopov}, \citenamefont {Malomed},\ and\ \citenamefont
  {Zeldovich}}]{Khlopov:1985jw}%
  \BibitemOpen
  \bibfield  {author} {\bibinfo {author} {\bibfnamefont {M.}~\bibnamefont
  {Khlopov}}, \bibinfo {author} {\bibfnamefont {B.~A.}\ \bibnamefont
  {Malomed}}, \ and\ \bibinfo {author} {\bibfnamefont {I.~B.}\ \bibnamefont
  {Zeldovich}},\ }\href@noop {} {\bibfield  {journal} {\bibinfo  {journal}
  {Mon. Not. Roy. Astron. Soc.}\ }\textbf {\bibinfo {volume} {215}},\ \bibinfo
  {pages} {575} (\bibinfo {year} {1985})}\BibitemShut {NoStop}%
\bibitem [{\citenamefont {Easther}\ \emph {et~al.}(2011)\citenamefont
  {Easther}, \citenamefont {Flauger},\ and\ \citenamefont
  {Gilmore}}]{Easther:2010mr}%
  \BibitemOpen
  \bibfield  {author} {\bibinfo {author} {\bibfnamefont {R.}~\bibnamefont
  {Easther}}, \bibinfo {author} {\bibfnamefont {R.}~\bibnamefont {Flauger}}, \
  and\ \bibinfo {author} {\bibfnamefont {J.~B.}\ \bibnamefont {Gilmore}},\
  }\href {\doibase 10.1088/1475-7516/2011/04/027} {\bibfield  {journal}
  {\bibinfo  {journal} {JCAP}\ }\textbf {\bibinfo {volume} {1104}},\ \bibinfo
  {pages} {027} (\bibinfo {year} {2011})},\ \Eprint
  {http://arxiv.org/abs/1003.3011} {arXiv:1003.3011 [astro-ph.CO]} \BibitemShut
  {NoStop}%
\bibitem [{\citenamefont {Berges}(2015)}]{Berges:2015kfa}%
  \BibitemOpen
  \bibfield  {author} {\bibinfo {author} {\bibfnamefont {J.}~\bibnamefont
  {Berges}},\ }\href@noop {} {\  (\bibinfo {year} {2015})},\ \Eprint
  {http://arxiv.org/abs/1503.02907} {arXiv:1503.02907 [hep-ph]} \BibitemShut
  {NoStop}%
\end{thebibliography}%

\end{document}